\documentclass[12pt]{iopart}

\usepackage{graphicx}
\usepackage{amssymb}
\usepackage{iopams}
\usepackage{float}

\def\beq{\begin{equation}}
\def\eeq{\end{equation}}
\def\br{\begin{eqnarray}}
\def\er{\end{eqnarray}}
\def\benu{\begin{enumerate}}
\def\eenu{\end{enumerate}}

\def\l{\left}
\def\r{\right}

\begin{document}
\title[Punctuated inflation and the low CMB multipoles]{Punctuated 
inflation and the low CMB multipoles}
\author{Rajeev Kumar Jain$^{1+}$, Pravabati Chingangbam$^{2\dag}$,\\
Jinn-Ouk Gong$^{3\ast}$, L.~Sriramkumar$^{1\star}$
and Tarun Souradeep$^{4\S}$}
\address{$^{1}$Harish-Chandra Research Institute, Chhatnag Road,
Jhunsi,\\
${}$~Allahabad~211~019, India.}
\address{$^{2}$Korea Institute for Advanced Study,
207--43 Cheongnyangni 2-dong,\\ 
${}$~Dongdaemun-gu, Seoul 130-722, Korea.}
\address{$^{3}$Department of Physics, University of Wisconsin-Madison,\\
${}$~1150 University Avenue, Madison, WI 53706-1390, U.S.A.}
\address{$^{4}$IUCAA, Post Bag 4, Ganeshkhind, Pune 411 007, India.}
\eads{\mailto{$^{+}$rajeev@hri.res.in}, \mailto{$^{\dag}$prava@kias.re.kr}, 
\mailto{$^{\ast}$jgong@hep.wisc.edu},\\ 
{\hskip 13mm}\mailto{$^{\star}$sriram@hri.res.in},
\mailto{$^{\S}$tarun@iucaa.ernet.in}}
\begin{abstract}
We investigate inflationary scenarios driven by a class of potentials 
which are similar in form to those that arise in certain minimal 
supersymmetric extensions of the standard model. 
We find that these potentials allow a brief period of departure from
inflation sandwiched between two stages of slow roll inflation. 
We show that such a background behavior leads to a step like feature
in the scalar power spectrum.
We set the scales such that the drop in the power spectrum occurs at a 
length scale that corresponds to the Hubble radius today---a feature that 
seems necessary to explain the lower power observed in the quadrupole 
moment of the Cosmic Microwave Background (CMB) anisotropies. 
We perform a Markov Chain Monte Carlo analysis to determine the values 
of the model parameters that provide the best fit to the recent WMAP 
$5$-year data for the CMB angular power spectrum. 
We find that an inflationary spectrum with a suppression of power at 
large scales that we obtain leads to a much better fit (with just
one extra parameter, $\chi_{\rm eff}^{2}$ improves by $6.62$) of the 
observed data when compared to the best fit reference $\Lambda$CDM 
model with a featureless, power law, primordial spectrum. 
\end{abstract}
\maketitle


\section{Introduction}

Measurements of the Cosmic Microwave Background (CMB) 
anisotropies---from the early days of the COsmic Background Explorer 
(COBE) satellite until the most recent observations of the Wilkinson 
Microwave Anisotropy Probe (WMAP)---have consistently indicated 
a low value of the quadrupole, below the cosmic variance of the 
concordant $\Lambda$CDM cosmological model with a nearly scale 
invariant, primordial 
spectrum~\cite{cobe-2,cobe-4,wmap-1,wmap-3,wmap-5}.
While there has been a recurring debate about the statistical 
significance of the quadrupole and the other outliers (notably,
near the multipole moments $22$ and $40$) in the 
CMB angular power spectrum (see 
Refs.~\cite{odwyer-2004,magueijo-2007,park-2007,chiang-2007} and
references therein), there has also been constant activity to 
understand possible underlying physical reasons for the outliers 
(see, for an inexhaustive list,
Refs.~\cite{starobinsky-1992,jing-1994,feng-2003,efsthathiou-2003,luminet-2003,hajian-2003-05,kawasaki-2003-05,gordon-2004,contaldi-2003,cline-2003,moroi-2004,hunt-2004-07,sriram-2004a,sriram-2004b,sriram-2005,sinha-2006,covi-2006-07,boyanovsky-2006,donoghue-2007,powell-2007,nicholson-2008,joy-2007-08,destri-2008}).

Given the CMB observations, different model independent approaches 
have been used to recover the primordial spectrum (see, for example,
Refs.~\cite{wmap-3,bridle-2003,mukherjee-2003,hannestad-2004,tarun-2004-08,tv-2005}).
While all these approaches arrive at a spectrum that is nearly scale 
invariant at the smaller scales, most of them inevitably seem to point 
to a sharp drop in power at the scales corresponding to the Hubble scale 
today.
Within the inflationary scenario, a variety of single and two field 
models have been constructed to produce such a drop in power at the 
large scales~\cite{starobinsky-1992,feng-2003,contaldi-2003,cline-2003,moroi-2004,boyanovsky-2006,powell-2007,nicholson-2008,destri-2008}.
However, in single field inflationary models, in order to produce 
such a spectrum, we find that many of the scenarios either assume 
a specific pre-inflationary regime, say, a radiation dominated
epoch, or special initial conditions for the background scalar
field, such as an initial period of fast roll~\cite{contaldi-2003,cline-2003,powell-2007,nicholson-2008}. 
Moreover, some of them impose the initial conditions on the 
perturbations when the largest scales are outside the Hubble 
radius during the pre-inflationary or the fast roll 
regime~\cite{contaldi-2003,powell-2007,nicholson-2008}. 
Such requirements are rather artificial and, ideally, it would be 
preferable to produce the desired power spectrum during an inflationary 
epoch without invoking any specific pre-inflationary phase or special 
initial conditions for the inflaton.
Furthermore, though a very specific pre-inflationary phase such 
as the radiation dominated epoch may allow what can be considered 
as natural (i.e. Minkowski-like) initial conditions for the 
perturbations even at super-Hubble scales, we believe that 
choosing to impose initial conditions for a small subset of 
modes when they are outside the Hubble radius, while demanding 
that such conditions be imposed on the rest of the modes at 
sub-Hubble scales, can be considered unsatisfactory.

It has long been known that power spectra with large deviations 
from scale invariance can be generated in inflationary models 
that admit one or more periods of departure from the slow roll 
phase (see, for instance, 
Refs.~\cite{starobinsky-1992,hodges-1990,mukhanov-1991,tarun-1995,adams-1997,lesgourgues-2000,barriga-2001,adams-2001,jinn-ouk-2005}). 
The degree of the deviation from a nearly scale invariant spectrum 
would be determined by the extent and duration of the departure, 
which are, in turn, controlled by the parameters of the model. 
A departure from slow roll affects the evolution of modes that leave 
the Hubble radius just before the departure. 
Rather than remaining constant, the curvature perturbations, say, 
${\cal R}_{k}$, corresponding to these modes evolve at super-Hubble 
scales, sourced by the intrinsic entropy perturbations of the inflaton 
field which, typically, exhibit a rapid growth during the 
fast roll regime~\cite{leach-2001,rajeev-2007}. 
Such an evolution on super-Hubble scales results in dips or bursts 
of oscillations in the scalar power spectrum. 
Usually, such a departure is induced by introducing a sharp feature in
the potential of the inflaton field, such as a step or a sudden change in 
the slope~\cite{starobinsky-1992,covi-2006-07,joy-2007-08,jinn-ouk-2005}. 
However, this is not necessary, and transitions to fast roll for 
brief periods can be generated even with smooth and better motivated 
effective potentials~\cite{hodges-1990,tarun-1995,leach-2001,rajeev-2007}.

Our purpose in this paper is to present a simple model of inflation that supresses the power spectrum on large scales, a feature---as we discussed
above---that seems to be necessary to fit the lower power in the quadrupole 
(and, to some extent, in the deviant power at other lower multipoles 
such as the octopole and the multipole $\ell=22$) of the CMB angular 
power spectrum, using an effective potential of the canonical scalar 
field {\it without}\/ introducing any ad hoc sharp feature. 
We find that the form of the potentials motivated by a class of certain 
minimal supersymmetric extensions of the standard model provide us with 
the desired behavior~\cite{allahverdi-2006,allahverdi-2007,sanchez-2007,allahverdi-2008}. 
These large field models allow a period of fast roll sandwiched between 
two stages of slow roll inflation\footnote{Earlier, in the literature, 
two successive stages of slow roll inflation have often been driven by 
two scalar fields~\cite{kofman-1985,silk-1987,polarski-1992-95,langlois-1999,tsujikawa-2003}.  
Instead, in this paper, we achieve the two stages of slow roll inflation 
including a brief period of departure from inflation, all with just a 
single scalar field.}. 
The first phase of slow roll inflation allows us to impose the 
standard Bunch-Davies initial conditions on the modes which exit 
the Hubble radius during the subsequent fast roll regime, an epoch
due to which the curvature perturbations on the super-Hubble scales 
are suppressed. 
The second slow roll phase lasts for about $50$-$60$ $e$-folds, thereby 
allowing us to overcome the standard horizon problem associated with
the hot big bang model. 
The advantages of our approach over other single field models mentioned 
earlier are twofold. 
Firstly, we do not need to assume any specific pre-inflationary phase. 
The entire evolution of the inflationary era is described by a single 
inflaton potential and, therefore, is much simpler. 
Secondly, the modes which exit the Hubble radius during the fast roll 
regime are inside the Hubble radius during the first stage of slow 
roll inflation and, hence, we do not have to impose any special initial
conditions on the large scale modes.

This paper is organized as follows.
In Sec.~\ref{sec:bd}, we shall review the essential features of the 
effective inflaton potential that we shall consider, and describe 
the background dynamics in situations of our interest.
In Sec.~\ref{sec:ps}, after an outline of the slow roll `expectations'
of the scalar spectrum that can arise in such a background, we shall 
discuss the spectra that we obtain through numerical integration.
In Sec.~\ref{sec:comparison}, using the cosmological Boltzmann code
CAMB and the Monte Carlo code COSMOMC, we shall compare the power 
spectra from the models we consider with the recent WMAP 5-year data.
Finally, we shall close with Sec.~\ref{sec:discussion}, wherein after 
a brief summary of our results, we shall discuss as to how the results 
from our model compare with those that have been obtained in another 
closely related single field model.

In the discussions below, we shall set $\hbar$ and $c$ as well 
as $M_{_\mathrm{Pl}} = (8\,\pi\, G)^{-1/2}$ to unity.
Also, throughout, an overdot and an overprime shall denote 
differentiation with respect to the cosmic and the conformal
times, respectively.


\section{The inflaton potential and the background dynamics}\label{sec:bd}

\begin{figure}[!htb]
\begin{center}
\vskip 20pt
\resizebox{220pt}{160pt}{\includegraphics{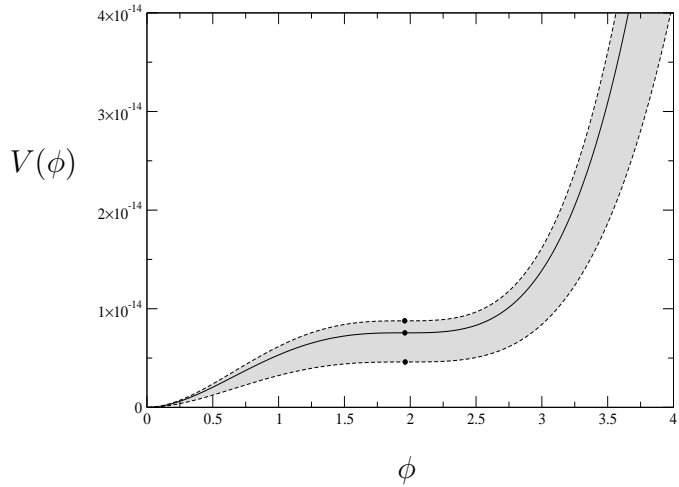}}
\vskip -112 true pt \hskip -260 true pt $V(\phi)$
\vskip 100 true pt \hskip 15 true pt $\phi$
\vskip 2 true pt
\end{center}
\caption{Illustration of the inflaton potential~(\ref{eq:mssm-p}) for
$n=3$. 
The solid line corresponds to the following values for the potential
parameters: $m=1.5368\times 10^{-7}$ and $\lambda=6.1517\times10^{-15}$
(corresponding to $\phi_{0}=1.9594$), values which turn out to provide 
the best fit to the WMAP $5$-year data (cf.~Tab.~\ref{tab:bfv}).
The dashed lines correspond to values that are $1$-$\sigma$ away from 
the best fit ones. 
The black dots denote the points of inflection.}
\label{fig:mssm-p}
\end{figure}

The effective potential for the inflation field that we shall consider
is described by two parameters $m$ and $\lambda$, and is given by
\beq
V(\phi) = \l(\frac{m^2}{2}\r)\,\phi^2  
- \l(\frac{\sqrt{2\,\lambda\,(n-1)}\,m}{n}\r)\, \phi^n 
+ \l(\frac{\lambda}{4}\r)\,\phi^{2(n-1)},
\label{eq:mssm-p}
\eeq
where $n>2$ is an integer. 
Such potentials are known to arise in certain minimal supersymmetric
extensions of the standard model~\cite{allahverdi-2006}, and their 
role as an inflaton and its related effects have been studied recently~\cite{allahverdi-2007,sanchez-2007,allahverdi-2008}. 
(We should also hasten to add that the specific case of $n=3$ has been
considered much earlier for reasons similar to ours, viz. producing
certain features in the primordial spectrum~\cite{hodges-1990}.)
In the above potential, the coefficient of the $\phi^n$ term has been 
chosen in such a way that the potential has a point of inflection at 
$\phi = \phi_0$ (i.e. the location where both $V_{\phi}\equiv 
(dV/d\phi)$ and $V_{\phi\phi}\equiv (d^{2}V/d\phi^{2})$ vanish), with 
$\phi_{0}$ given by
\beq
\phi_{0} 
= \left[\frac{2\,m^2}{(n-1)\,\lambda}\right]^{\frac{1}{2\,(n-2)}}.
\eeq
Near this point of inflection, the potential exhibits a plateau with an 
extremely small curvature, which, as we shall discuss below, proves to 
be crucial for the desired evolution of the inflaton field. 
The potential~(\ref{eq:mssm-p}) for the case $n=3$ is depicted in 
Fig.~\ref{fig:mssm-p}.

Note that the potential~(\ref{eq:mssm-p}) roughly behaves as
\begin{equation}
V(\phi) \sim \left\{
\begin{array}{ll}
\phi^{2(n-1)}, & \mbox{for} \quad \phi > \phi_{0},\\
\phi^2, & \mbox{for} \quad \phi < \phi_{0} .
\end{array}
\right.
\end{equation}
Recall that, the first potential slow roll parameter is given 
by~\cite{texts,reviews}
\beq
\epsilon_1 = \l(\frac{1}{2}\r)\, \l(\frac{V_\phi}{V}\r)^2,
\eeq
and inflation ends as $\epsilon_1$ crosses unity.
It is then clear that, in a power law potential of the form $V \sim
\phi^{2(n-1)}$, slow roll inflation will occur (i.e. $\epsilon_1 \ll 1$)
when $\phi \gg 1$, and inflation will end when $\phi_\mathrm{end} 
\simeq \l[\sqrt{2}\, (n-1)\r] \sim \mathcal{O}(1)$. 
Thus, for a transition from slow roll to fast roll to occur, we need 
to choose the two parameters in the potential~(\ref{eq:mssm-p}) so 
that $\phi_0 \sim \mathcal{O}(1)$, i.e. of the order of the (reduced) 
Planck scale.
Restarting inflation after the fast roll phase and the number of 
$e$-folds that can be achieved during the second phase of slow 
roll crucially depends on the value of $\phi_0$. 
We rely on the numerics to choose this parameter carefully since the 
above slow roll estimate only provides a rough order of magnitude. 
Choosing $\phi_0$ in such a way is actually fine tuning, but it 
seems to be inevitable if we are to achieve the desired 
slow-fast-slow roll transition as well as the required number of 
$e$-folds. 
Once the point of inflection has been identified, we find that 
the normalization to the CMB angular power spectrum data provides 
the second constraint, thereby determining the value of the other 
free parameter~$m$.

\begin{figure}[!htb]
\begin{center}
\resizebox{300pt}{180pt}{\includegraphics{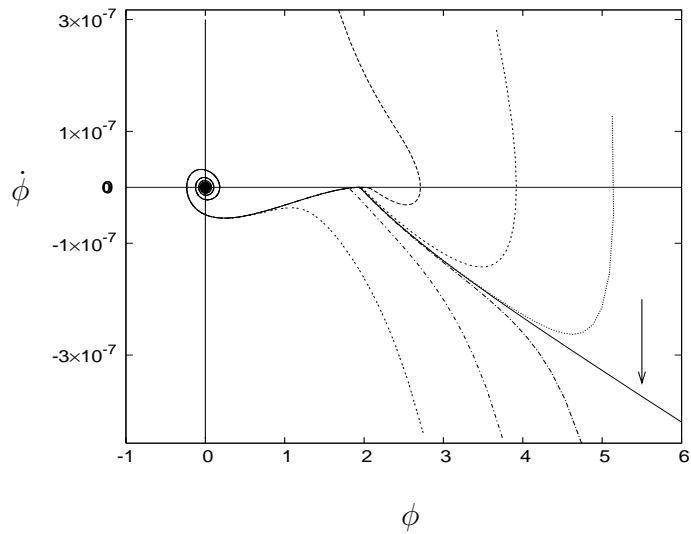}}
\vskip -118 true pt \hskip -220 true pt $\dot\phi$
\vskip 107 true pt \hskip 75 true pt $\phi$
\vskip 2 true pt
\end{center}
\caption{The phase portrait of the scalar field described by the
potential~(\ref{eq:mssm-p}) in the case of $n=3$ and for the values 
of the parameters $m$ and $\lambda$ mentioned in the last figure.
The arrow points to the attractor. 
Note that, as discussed in the text, all the trajectories quickly 
approach the attractor. 
We should mention that, though we have plotted the phase portrait 
for just the $n=3$ case, we find that such a behavior is exhibited 
by higher values of $n$ (such as, for example, $n=4,6$) as well.}
\label{fig:pp}
\end{figure}

The equation of motion governing the scalar field described by the
potential~(\ref{eq:mssm-p}), when expressed as two first order 
equations for the coupled variables $\phi$ and $\dot\phi$, has one 
attractive fixed point located at the origin, i.e. at $(\phi, {\dot
\phi})=(0,0)$. 
For positive values of~$\phi$, we find that there exists an attractor 
trajectory towards which all other trajectories with arbitrary 
initial conditions on $\phi$ and $\dot\phi$ quickly converge. 
For a suitably chosen $\phi_0$, we find that the attractor trajectory 
exhibits two regimes of slow roll inflation sandwiching a period of 
fast roll. 
Hence, if we start the evolution with $\phi \gg \phi_0$, then the 
initial values of $\phi$ and ${\dot \phi}$ prove to be irrelevant 
for the subsequent dynamics as they approach the attractor.  
This behavior is evident from Fig.~\ref{fig:pp} where we have plotted 
the phase portrait for the $n = 3$ case.
Once the field reaches close to $\phi_0$, due to the extreme flatness 
of the potential~(\ref{eq:mssm-p}), it relaxes and then moves very 
slowly, commencing the second stage of the slow roll inflation. 
This stage ends when the field finally rolls down towards the minima 
of the potential at $\phi=0$. 

We had mentioned earlier that potentials of the type~(\ref{eq:mssm-p}) 
are encountered in the Minimal Supersymmetric Standard Model
(MSSM)~\cite{allahverdi-2006}, and that their role as an inflaton
has been analyzed recently~\cite{allahverdi-2007,sanchez-2007,allahverdi-2008}. 
At this point it is important that we highlight the differences between 
MSSM inflation and the scenario we are considering. 
In MSSM inflation, the point of inflection is located at sub-Planckian
values (i.e. $\phi_0 \ll 1$) which can be an advantage as it avoids the 
problems associated with having super-Planckian values for the field.  
In contrast, in our case, as emphasized above, the saddle point should 
be located around the Planck scale (i.e. $\phi_0 \gtrsim 1$), if we are 
to achieve the second period of slow roll before the end of inflation.
However, in the MSSM case, to have successful inflation, the initial 
values of $\phi$ and ${\dot \phi}$ have to be finely tuned so that 
$\phi_{\rm ini} \simeq \phi_0$ and ${\dot \phi}_{\rm ini}\simeq 0$. 
But, in our scenario, we do not require such fine tuning of the initial conditions on $\phi$ and ${\dot \phi}$. 
Instead, we require for the location of $\phi_0$.    

Though the parameters of the potential that we work with are different 
from the MSSM case, we nevertheless believe that it may be possible to
realize the potential~(\ref{eq:mssm-p}) in theories beyond the standard
model, such as, for instance, string theory  (in this context, see, for
example, Refs.~\cite{itzhaki-2007,ashoorioon-2006-08}). 
For example, it is known that the existence of a number of string axion 
fields can give rise to the following potential describing multi-field 
chaotic inflation~\cite{dimopoulos-2008}:
\beq
V(\phi_{i}) = \sum_i \l(\frac{1}{2}\r)\, m_i^{2}\, \phi_i^2,
\eeq
with the initial field displacements smaller than unity. 
The dynamics and the inflationary predictions in such examples are 
surprisingly similar to the corresponding single field chaotic 
inflation models~\cite{alabidi-2005,kim-2006,gong-2007}, due to the 
assisted inflation mechanism~\cite{liddle-1998}. 
Similarly, with enough number of fields and with the non-renormalizable 
superpotential
\beq
W = \l(\frac{\lambda}{n}\r)\;\l(\frac{\phi^n}{M_{_\mathrm{Pl}}^{n-3}}\r),
\label{eq:sp}
\eeq
and the corresponding $A$ term and the soft mass term, one might be able 
to build an inflation model that is effectively equivalent to the single 
field one described by the potential~(\ref{eq:mssm-p}).
(Note that, for clarity, we have temporarily restored $M_{_\mathrm{Pl}}$
in the expression~(\ref{eq:sp}) above.)


\section{The scalar power spectrum}\label{sec:ps}

In this section, after providing general arguments for the form of
the scalar spectra that we can expect from our model, we present 
the spectra evaluated numerically.


\subsection{Key equations and essential quantities}

Let us begin by quickly summarizing the essential equations and the 
quantities that we are interested in~\cite{texts,reviews}.
The curvature perturbation ${\cal R}_{k}$ satisfies the differential 
equation
\beq
{\cal R}_{k}''+2\, \l(\frac{z'}{z}\r)\, {\cal R}_{k}'
+k^{2}\, {\cal R}_k=0,\label{eq:deRk}
\eeq
where the quantity $z$ is given by
\beq
z = \l(a\,\phi'/{\cal H}\r).
\eeq
The quantity $a$ denotes the scale factor, $\phi$ the background 
inflaton, and ${\cal H}$ is the conformal Hubble factor given by 
$(a'/a)$.
The scalar power spectrum ${\cal P}_{_{\rm S}}(k)$ is then defined as
\beq
{\cal P}_{_{\rm S}}(k) = \l(\frac{k^{3}}{2\, \pi^{2}}\r)\,
\vert{\cal R}_{\rm k}\vert^{2},
\eeq
with the amplitude of the curvature perturbation ${\cal R}_{k}$ evaluated,
in general, in the super-Hubble limit.
The tensor perturbation ${\cal U}_{k}$ satisfies the equation
\beq
{\cal U}_{k}''+2\, \l(\frac{a'}{a}\r)\, {\cal U}_{k}'
+k^{2}\; {\cal U}_k=0,\label{eq:deuk}
\eeq
with the tensor power spectrum ${\cal P}_{_{\rm T}}(k)$ being given by
\beq
{\cal P}_{_{\rm T}}(k) 
= \l(\frac{k^{3}}{2\pi^{2}}\r)\,
\vert{\cal U}_{\rm k}\vert^{2},
\eeq
where, as in the scalar case, the tensor amplitude ${\cal U}_{k}$ is
evaluated at super-Hubble scales. 
Finally, the tensor-to-scalar ratio $r$ is defined as follows:
\beq
r \equiv \l(\frac{{\cal P}_{_{\rm T}}} {{\cal P}_{_{\rm S}}}\r).
\eeq


\subsection{Physical `expectations'}


Before we evaluate the scalar spectra numerically, let us broadly 
try and understand the spectra that we can expect to arise in the 
slow-fast-slow roll scenario that we are interested in.


\subsubsection{The evolution of the scalar modes and the scalar spectrum}

Consider modes that exit the Hubble scale during an epoch of slow roll 
inflation.
Provided there is no deviation from slow roll soon after the modes 
leave the Hubble radius, the amplitude of these modes will remain 
constant at super-Hubble scales.
Therefore, their amplitude is determined by their value at Hubble exit, 
and the scalar power spectrum corresponding to these modes can be 
expressed in terms of the potential as follows~\cite{texts,reviews}:
\beq
{\cal P}_{_{\rm S}}(k) 
\simeq \l(\frac{1}{12\,\pi^2}\r)\, 
\l(\frac{V^3}{V_{\phi}^2}\r).
\eeq
However, if there is a period of deviation from slow roll 
inflation, then the asymptotic (i.e. the extreme super-Hubble) 
amplitude of the modes that leave the Hubble radius {\it just 
before}\/ the deviation are enhanced when compared to their 
value at Hubble exit~\cite{leach-2001}.
While modes that leave well before the deviation remain unaffected, 
it is found that there exists an intermediate range of modes whose 
amplitudes are actually {\it suppressed}\/ at super-Hubble 
scales~\cite{rajeev-2007}. 
As a result, in the slow-fast-slow roll scenario of our interest,
the scalar power spectrum is initially characterized by a sharp 
dip and a rise corresponding to modes that leave the Hubble radius 
just before the transition to fast roll.
Then arises a regime of nearly scale invariant spectrum corresponding 
to modes that leave during the second stage of slow roll inflation. 


\subsubsection{The effects on the tensor modes and the tensor spectrum}

Let us now understand the behavior of the tensor modes.
In the case of the scalar modes, the quantity $(z'/z)$ that appears 
in the differential equation~(\ref{eq:deRk}) turns out to be negative 
during a period of fast roll, and it is this feature that proves to be 
responsible for the amplification or the suppression of the modes at 
super-Hubble scales~\cite{leach-2001,rajeev-2007}. 
In contrast, the coefficient of the friction term in the 
equation~(\ref{eq:deuk}) that describes the tensor modes, 
viz. $(2\, {\cal H})$, is a positive definite quantity.
Hence, we do not expect any non-trivial super-Hubble evolution 
of ${\cal U}_{k}$.
We find that, in the models that we consider, the tensor-to-scalar 
ratio $r$ remains smaller than $10^{-4}$ over scales of cosmological 
interest, which is below the levels of possible detection by 
forthcoming missions such as PLANCK~\cite{planck}.


\subsection{Numerical results}

\begin{figure}[!htb]
\begin{center}
\vskip 25pt
\resizebox{270pt}{180pt}{\includegraphics{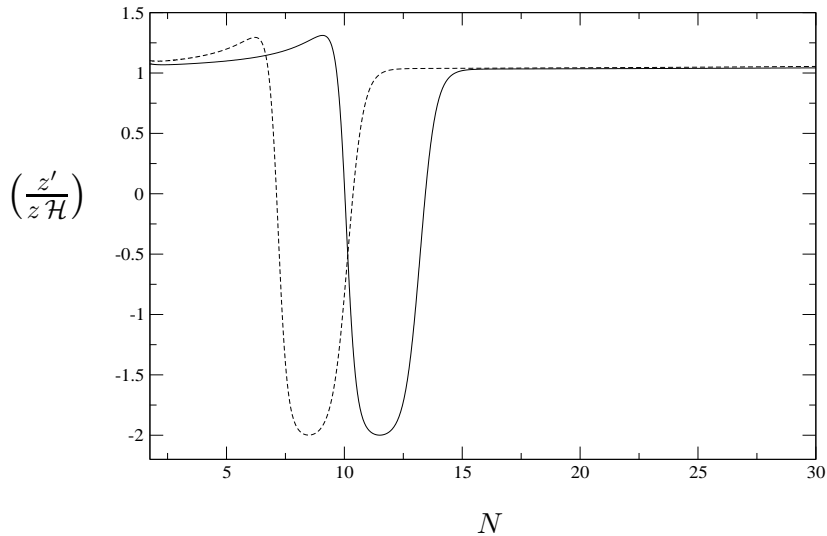}}
\vskip -122 true pt \hskip -320 true pt 
\large{$\l(\frac{z'}{z\, {\cal H}}\r)$}
\vskip 110 true pt \hskip 15 true pt \small{$N$}
\vskip 2 true pt
\caption{The background quantity $(z'/z\, {\cal H})$ has been 
plotted as a function of the number of $e$-folds, say, $N$, for 
the cases of $n=3$ and $n=4$ in potential~(\ref{eq:mssm-p}).
The solid line represents the $n=3$ case with the same values for 
the potential parameters as in the previous two figures.
The dashed line corresponds to the $n=4$ case with $m =1.1406\times 
10^{-7}$ and $\lambda =1.448\times 10^{-16}$ (corresponding to 
$\phi_{0}=2.7818$) and, as in the $n=3$ case, we have chosen 
these values as they provide the best fit to the WMAP $5$-year 
data. 
Also, note that we have imposed the following initial conditions 
for the background field in both the cases: $\phi_{\rm ini}=10$ 
and ${\dot \phi}_{\rm ini}=0$.
Evidently, the $n=3$ case departs from slow roll when $7\lesssim N 
\lesssim 15$, while the departure occurs during $4\lesssim N \lesssim 
12$ in the case of $n=4$.} 
\label{fig:zpz}
\end{center}
\end{figure}

\begin{figure}[!htb]
\begin{center}
\vskip 25pt
\resizebox{270pt}{180pt}{\includegraphics{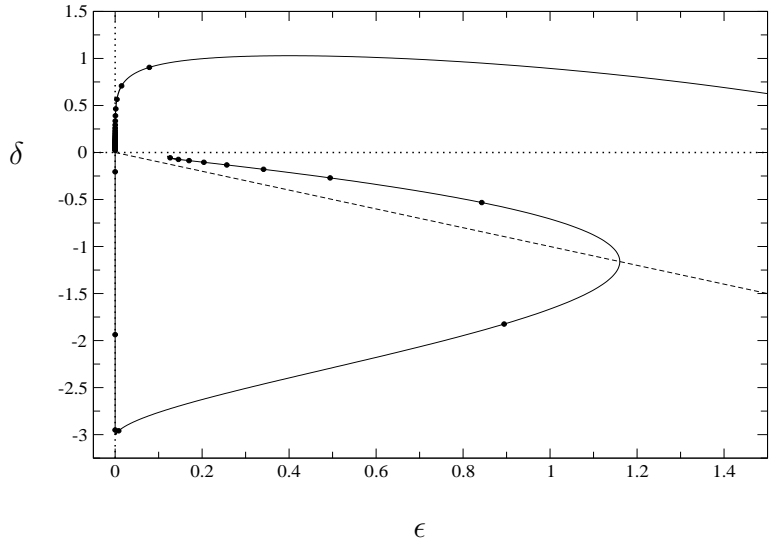}}
\vskip -135 true pt \hskip -300 true pt $\delta$
\vskip 125 true pt \hskip 5 true pt $\epsilon$
\vskip 2 true pt
\end{center}
\caption{The evolution of the scalar field has been plotted (as 
the solid black line) in the plane of the first two Hubble slow 
roll parameters $\epsilon$ and $\delta$ in the case of $n=3$ and 
for the best fit values of the parameters $m$ and $\lambda$ we 
have used earlier in Figs.~\ref{fig:mssm-p} and~\ref{fig:pp}.
The black dots have been marked at intervals of one $e$-fold, 
while the dashed line corresponds to $\epsilon=-\delta$.
Note that $\epsilon >1$ during $8<N<9$.
In other words, during fast roll, inflation is actually interrupted
for about a $e$-fold.}
\label{fig:ed}
\end{figure}

\begin{figure}[!htb]
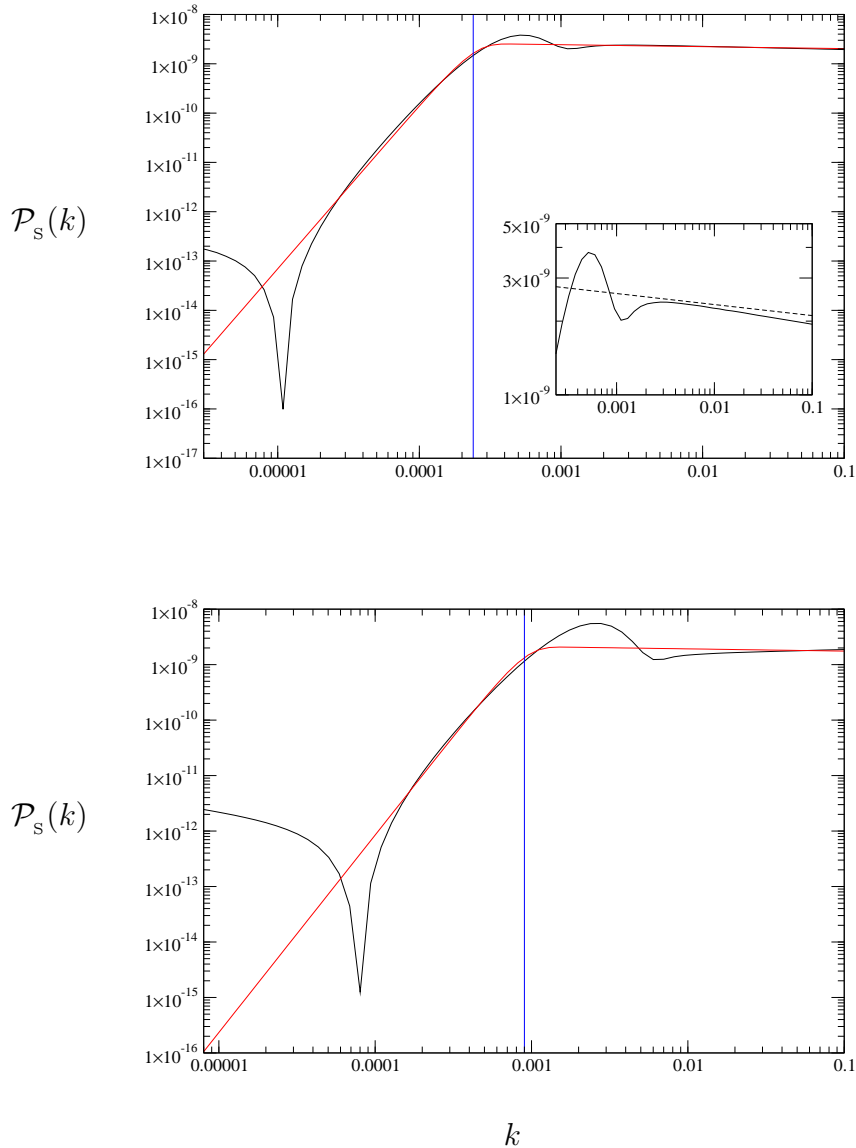

\begin{center}
\vskip 25pt\noindent
\resizebox{270pt}{180pt}{\includegraphics{ps-n3.eps}}
\vskip -110 true pt \hskip -340 true pt ${\cal P}_{_{\rm S}}(k)$ 
\vskip 135pt
\resizebox{270pt}{180pt}{\includegraphics{ps-n4.eps}}
\vskip -110 true pt \hskip -340 true pt ${\cal P}_{_{\rm S}}(k)$
\vskip 105 true pt \hskip 10 true pt $k$
\caption{The scalar power spectrum ${\cal P}_{_{\rm S}}(k)$ (the 
solid black line) have been plotted as a function of the 
wavenumber~$k$ for the cases of $n=3$ (on top) and $n=4$ (at the 
bottom).
We have chosen the same values for the potential parameters as in 
the earlier figures.
Moreover, we should emphasize that we have arrived at these spectra 
by imposing the standard, Bunch-Davies, initial condition on 
{\it all}\/ the modes.
The red line in these plots is the spectrum~(\ref{eq:ps-ecm}) with 
the exponential cut off.
It corresponds to $A_{_{\rm S}} = 2\times 10^{-9}$, $n_{_{\rm S}}\simeq 0.945$,
$\alpha=3.35$ and $k_{\ast}= 2.4\times 10^{-4}\; {\rm Mpc}^{-1}$ in the $n=3$ 
case, while $A_{_{\rm S}} = 2\times 10^{-9}$, $n_{_{\rm S}}\simeq 0.95$, 
$\alpha=3.6$ and $k_{\ast} = 9.0\times 10^{-4}\; {\rm Mpc}^{-1}$ in the case of 
$n=4$.
Note that the vertical blue line denotes~$k_{\ast}$.
The inset in the top panel illustrates the difference between our model
and the standard power law case (i.e. when ${\cal P}_{_{\rm S}}(k)=A_{_{\rm S}}\; 
k^{n_{_{\rm S}}-1}$, with the best fit values $A_{_{\rm S}}=2.1\times 10^{-9}$ 
and $n_{_{\rm S}}\simeq 0.955$) at smaller scales. 
This disparity leads to a difference in the CMB angular power spectrum 
at the higher multipoles, which we have highlighted in the inset in 
Fig.~\ref{fig:cl}.}
\label{fig:ps-n34}
\end{center}
\end{figure}

It is the background quantity $(z'/z)$ that appears in the 
differential equation~(\ref{eq:deRk}) for the curvature 
perturbation which essentially determines the form of the 
scalar power spectrum.
The quantity $(z'/z\, {\cal H})$ can be expressed in terms 
of the first two Hubble slow roll parameters, viz. $\epsilon
=-({\dot H}/H^2)$ and $\delta=({\ddot \phi}/H\, 
{\dot \phi})$, with $H=({\dot a}/a)$ being the standard 
Hubble parameter.
It is given by~\cite{leach-2001}
\beq
\l(\frac{z'}{z\, {\cal H}}\r) 
=\l(1+\epsilon+\delta\r),
\eeq
and it is clear from this expression that, during slow roll 
inflation (i.e. when $\epsilon \ll 1$ and $\delta \ll 1$), 
the quantity $(z'/z\, {\cal H})$ will remain close to 
unity~\cite{leach-2001,rajeev-2007}.
In Fig.~\ref{fig:zpz}, we have plotted the evolution of $(z'/z\, 
{\cal H})$ as a function of the number of $e$-folds $N$ for the 
cases of $n=3$ and $n=4$ in the potential~(\ref{eq:mssm-p}).
And, in Fig.~\ref{fig:ed}, we have plotted the evolution of the 
field in the plane of the Hubble slow roll parameters $\epsilon$ 
and $\delta$ for the $n=3$ case.
It is manifest from these figures that the departure from slow roll 
occurs roughly between $e$-folds $7 \lesssim N \lesssim 15$ in the 
$n = 3$ case and between $e$-folds $4 \lesssim N \lesssim 12$ for 
$n=4$.
We should also point out that inflation is actually interrupted for
about a $e$-fold during the fast roll.
In Fig.~\ref{fig:ps-n34}, we have plotted the corresponding scalar  
spectra evaluated numerically.
The broad arguments we had presented in the previous subsection are 
evidently corroborated by these two figures.
Note that, in plotting all these figures, we have chosen parameters that 
eventually provide the best fit to the WMAP $5$-year data.
Also, in the inset in the top panel of Fig.~\ref{fig:ps-n34}, we have 
highlighted the difference between the scalar spectra in our model and 
the power law case (i.e. when ${\cal P}_{_{\rm S}}(k)=A_{_{\rm S}}\; 
k^{n_{_{\rm S}}-1}$, with $A_{_{\rm S}}=2.1\times 10^{-9}$ and 
$n_{_{\rm S}}\simeq 0.955$).
Moreover, we should stress here that the standard sub-Hubble, Bunch-Davies, 
initial conditions have been imposed on {\it all}\/ the modes in 
arriving at these spectra.

The scalar power spectrum with a drop in power at large scales is 
often approximated by an expression with an exponential cut off of 
the following form~\cite{contaldi-2003,cline-2003,sinha-2006}:
\beq
{\cal P}_{_{\rm S}}(k)
=A_{_{\rm S}}\, \biggl(1-\exp{\l[-(k/k_{\ast})^{\alpha}\r]}\biggr)\,
k^{n_{_{\rm S}}-1}.\label{eq:ps-ecm}
\eeq
In Fig.~\ref{fig:ps-n34}, we have also plotted this expression for values 
of $A_{_{\rm S}}$, $n_{_{\rm S}}$, $\alpha$ and $k_{\ast}$ that closely 
approximate the spectra we obtain.
It is useful to note that the spectra we obtain correspond to 
$A_{_{\rm S}} = 2\times 10^{-9}$, $n_{_{\rm S}}\simeq 0.945$,
$\alpha=3.35$ and $k_{\ast} = 2.4\times 10^{-4}\; {\rm Mpc}^{-1}$ 
when $n=3$, while $A_{_{\rm S}} = 2\times 10^{-9}$, $n_{_{\rm S}}\simeq 0.95$,
$\alpha=3.6$ and $k_{\ast} = 9.0\times 10^{-4}\; 
{\rm Mpc}^{-1}$ in the $n=4$ case.
We should emphasize here that we have arrived at these values for $A_{_{\rm S}}$,
$n_{_{\rm S}}$, $\alpha$ and $k_{\ast}$ by a simple visual comparison of the
numerically evaluated result with the above exponentially cut off spectrum.


\section{Comparison with the recent WMAP $5$-year 
data}\label{sec:comparison}

In this section, we shall discuss as to how our model compares with 
the recent WMAP $5$-year data.


\subsection{The parameters in our model and the priors we work with}

In the standard concordant cosmological model---viz. the $\Lambda$CDM 
model with a power law inflationary perturbation spectrum---six 
parameters are introduced when comparing the theoretical results with 
the CMB data (see, for instance, Ref.~\cite{lewis-2008}).
Four of them are the following background parameters: the baryon 
density~$\l(\Omega_{\rm b}\, h^2\r)$, the density of cold dark 
matter~$\l(\Omega_{\rm c}\, h^2\r)$, the angular size of the 
acoustic horizon~$\theta$, and the optical depth $\tau$, with $h$ 
denoting the Hubble constant today (viz. $H_{0}$) expressed in 
units of $100\; {\rm km}\, {\rm s}^{-1}\, {\rm Mpc}^{-1}$.
The parameters that are introduced to describe the inflationary 
perturbation spectrum are the scalar amplitude~$A_{_{\rm S}}$ and 
the scalar spectral index~$n_{_{\rm S}}$.
The tensor-to-scalar ratio~$r$ is also introduced as a parameter 
provided the ratio is sufficiently large, say, when $r \gtrsim 
{\cal O} \l(10^{-2}\r)$.
However, in the models we consider, the tensor-to-scalar ratio 
proves to be smaller than $10^{-4}$ over the scales of cosmological 
interest.
So, we completely ignore the contribution due to the gravitational waves 
in our analysis.
We retain the standard background cosmological parameters, and we 
introduce the following three parameters to describe the inflationary 
perturbation spectrum: $m$, $\phi_{0}$ and $a_{0}$. 
While $m$ appears explicitly in the potential~(\ref{eq:mssm-p}), 
$\phi_{0}$ has been chosen in place of $\lambda$. 
The quantity $a_{0}$ denotes the initial value of the scale factor 
(i.e. at $N=0$), and it basically determines the location of the 
cut-off in the power spectrum.
Thus, we have one additional parameter in comparison with the standard 
case.
Essentially, we have traded off the scalar amplitude $A_{_{\rm S}}$ 
for $m$, and the scalar spectral index $n_{_{\rm S}}$ for $\phi_{0}$.
In Tab.~\ref{tab:priors}, we have listed the ranges of uniform priors 
that we have imposed on the various parameters.
\begin{table}[!htb]
\begin{center}
\begin{tabular}{|c|c|c|c|}
\hline\hline
Model & Parameter & Lower limit & Upper limit\\
\hline\hline
&$\Omega_{\rm b}\, h^2$ & $0.005$ & $0.1$\\
\cline{2-4}
Common &$\Omega_{\rm c}\, h^2$ & $0.001$ & $0.99$\\
\cline{2-4}
parameters &$\theta$ & $0.5$   & $10.0$\\
\cline{2-4}
&$\tau$ & $0.01$  & $0.8$\\
\hline
Reference & ${\rm log}\, \l[10^{10}\, 
A_{_{\rm S}}\r]$  & $2.7$   & $4.0$\\ 
\cline{2-4}
model&$n_{_{\rm S}}$ & $0.5$ & $1.5$\\
\hline
&${\rm log}\, \l[10^{10}\, m^2\r]$ & $-9.0$   & $-8.0$\\
\cline{2-4}
Our model&$\phi_{0}$ & $1.7$   & $2.3$\\
\cline{2-4}
&$a_{0}$ & $0.1$   & $2.0$\\
\hline\hline
\end{tabular}
\caption{The priors on the various parameters describing the reference 
$\Lambda$CDM model with a power law primordial spectrum and our model.
While the first four background cosmological parameters are common for
both the models, the fifth and the sixth parameters describe the power 
law primordial spectrum of the reference model.
As discussed in the text, in our model, we have traded off the 
scalar amplitude $A_{_{\rm S}}$ for $m$ and the spectral index 
$n_{_{\rm S}}$ for $\phi_{0}$.
The additional parameter in our model, viz. $a_{0}$, represents the 
value of the scale factor at $N=0$ and it essentially identifies the 
location of the cut-off in the power spectrum.}
\end{center}
\label{tab:priors}
\end{table}


\subsection{The best fit values and the joint constraints}

\begin{figure}
\begin{center}
${}$\hskip -15pt
\resizebox{400pt}{480pt}{\includegraphics{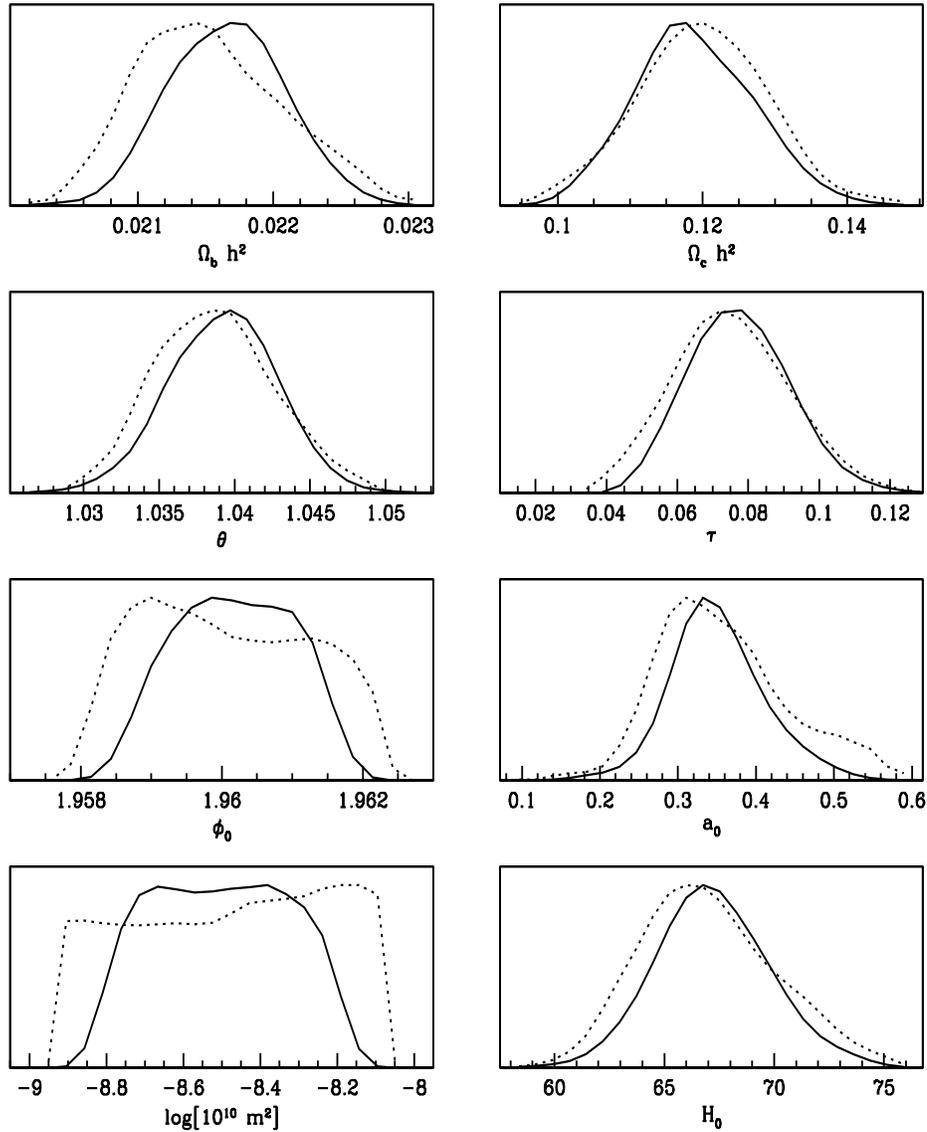}}
\end{center}
\vskip -30pt
\caption{The one-dimensional mean (the solid lines) and marginalized 
(dashed lines) likelihood curves for all the input parameters (and the
derived parameter $H_{0}$) in the $n=3$ case.}
\label{fig:lh-n3}
\end{figure}

\begin{figure}
\begin{center}
\resizebox{400pt}{480pt}{\includegraphics{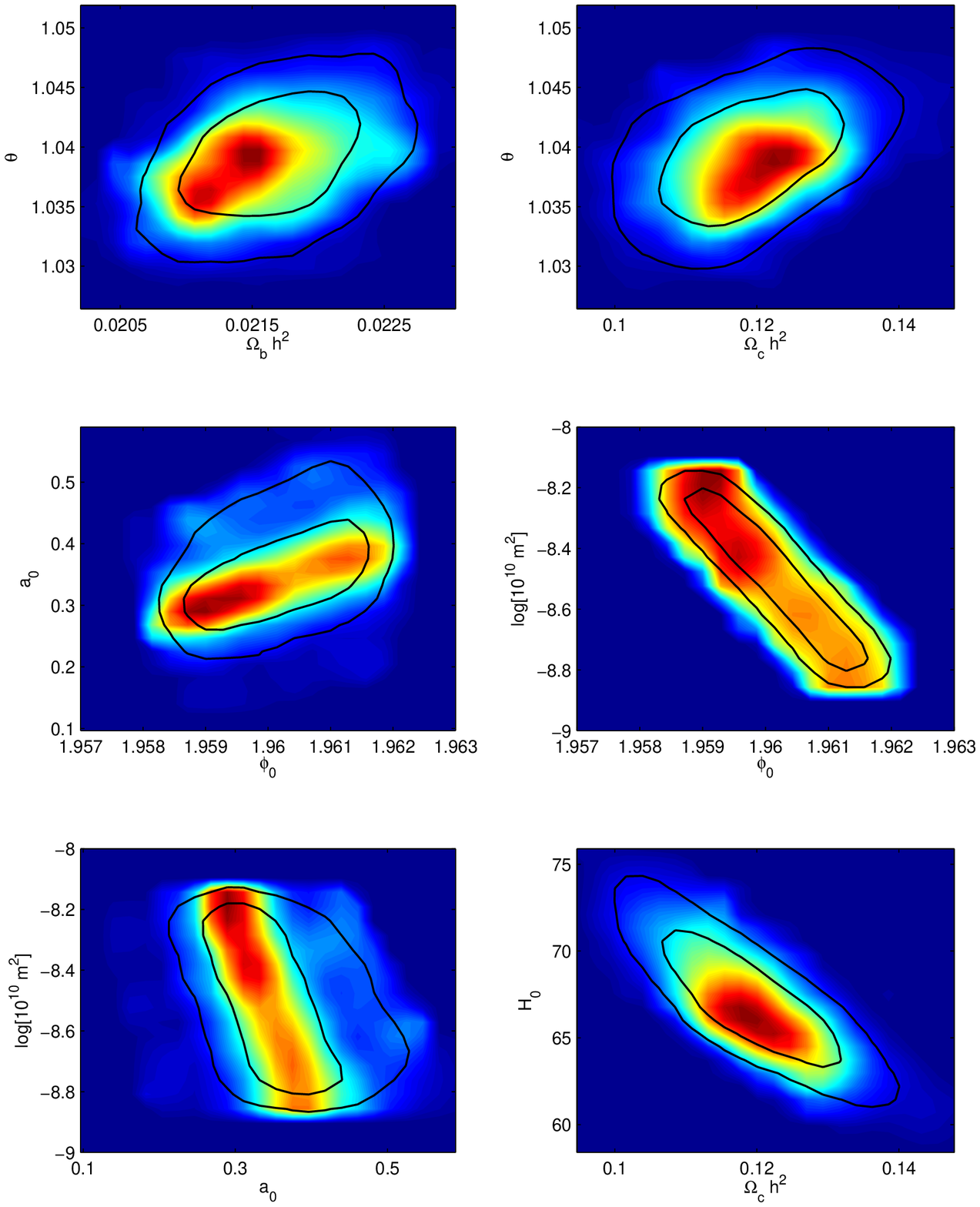}}
\end{center}
\caption{The $1$-$\sigma$ and $2$-$\sigma$ two-dimensional joint 
constraints on the different input parameters (and the derived 
parameter $H_{0}$) in the $n=3$ case.}
\label{fig:2dc3}
\end{figure}

We have compared the power spectra for the $n=3$ and the $n=4$ cases 
with the recent WMAP $5$-year data for the temperature-temperature,
the temperature-electric polarization and the electric-electric
polarization angular power spectra of the CMB anisotropies~\cite{wmap-5}.
We have used a modified version of the cosmological Boltzmann code 
CAMB~\cite{lewis-2000,camb} to calculate the angular power spectra 
of the CMB anisotropies, with the inflationary perturbation spectrum 
computed from a separate routine. 
We have evaluated the likelihood function using the likelihood code 
that has been made publicly available by the WMAP team~\cite{lambda}.
We have obtained the best fit values for the parameters of our model 
using COSMOMC~\cite{lewis-2002,cosmomc}, the publicly available, Markov 
Chain Monte Carlo (MCMC) code for the parameter estimation of a 
given cosmological model.
The MCMC convergence diagnostics are done on multiple parallel chains 
using the Gelman and Rubin (``variance of chain means''/``mean of chain variances'') $R$ statistics for each parameter, demanding that $(R-1) 
< 0.01$, a procedure that essentially looks at the fluctuations amongst 
the different chains and decides when to terminate the run.
We find that while the $n=3$ case provides a better fit to the data
than the reference concordant model~\cite{lewis-2008}, the $n=4$ case 
leads to such a poor fit to the data that we do not consider it 
hereafter.
We attribute the poor fit by the $n=4$ case (and also in the cases
wherein $n>4$) to the large bump in the scalar power spectrum that
arises just before the spectrum turns scale invariant (cf. 
Fig.~\ref{fig:ps-n34}).
We have plotted the one-dimensional marginalized and mean 
likelihood curves for the various parameters in the $n=3$ 
case in Fig.~\ref{fig:lh-n3}.
And, in Fig.~\ref{fig:2dc3}, we have plotted the corresponding
$1$-$\sigma$ and $2$-$\sigma$ two-dimensional joint constraints 
on the various parameters.
We have listed the best fit values and the $1$-$\sigma$ constraints on 
the various parameters describing the reference model and the $n=3$
case in Tab.~\ref{tab:bfv}.
We find that the $n=3$ case provides a much better fit to the data 
than the reference model with an improvement in $\chi_{\rm eff}^{2}$ 
of $6.62$.
\begin{table}[!htb]
\begin{center}
\begin{tabular}{|c|c|c|}
\hline\hline
Parameter & Reference model & Our model\\
\hline
$\Omega_{\rm b}\, h^2$ & $0.02242^{+0.00155}_{-0.00127}$  
& $0.02146^{+0.00142}_{-0.00108}$\\
\hline
$\Omega_{\rm c}\, h^2$ & $0.1075^{+0.0169}_{-0.0126}$
& $0.12051^{+0.02311}_{-0.02387}$ \\
\hline
$\theta$ & $1.0395^{+0.0075}_{-0.0076}$ 
& $1.03877^{+0.00979}_{-0.00931}$\\
\hline
$\tau$ &  $0.08695^{+0.04375}_{-0.03923}$ 
& $0.07220^{+0.04264}_{-0.02201}$\\
\hline    
${\rm log}\, \l[10^{10}\, A_{_{\rm S}}\r]$ 
& $3.0456^{+0.1093}_{-0.1073}$ 
& ---\\
\hline
$n_{_{\rm S}}$ & $0.9555^{+0.0394}_{-0.0305}$  
& ---\\
\hline    
${\rm log}\, \l[10^{10}\, m^2\r]$ 
& --- & $-8.3509^{+0.1509}_{-0.1473}$\\
\hline
$\phi_{0}$ & --- & $1.9594^{+0.00290}_{-0.00096}$\\
\hline
$a_{0}$ & --- & $0.31439^{+0.02599}_{-0.02105}$\\
\hline\hline
\end{tabular}
\caption{The mean values and the $1$-$\sigma$ constraints on the 
various parameters that describe the reference model and our model.
As we mentioned in the text, we find that the $n=3$ case provides 
a much better fit to the data than the reference model with an 
improvement in $\chi_{\rm eff}^{2}$ of $6.62$.}
\end{center}
\label{tab:bfv}
\end{table}
It is clear from Figs.~\ref{fig:lh-n3} and~\ref{fig:2dc3} that the
constraint on the parameter $m$ is prior dominated.
In our model, it is the parameter~$m$ that determines the amplitude 
of the power spectrum when it is nearly scale invariant.
This amplitude, in turn, is essentially determined by the first 
peak of the CMB angular power spectrum.
We should mention here that our choice of priors for the 
parameter~$m$ has been arrived at by a simple visual fit 
to the first peak.


\subsection{The CMB angular power spectra for the best fit values}

\begin{figure}
\begin{center}
\vskip 25pt
\resizebox{300pt}{200pt}{\includegraphics{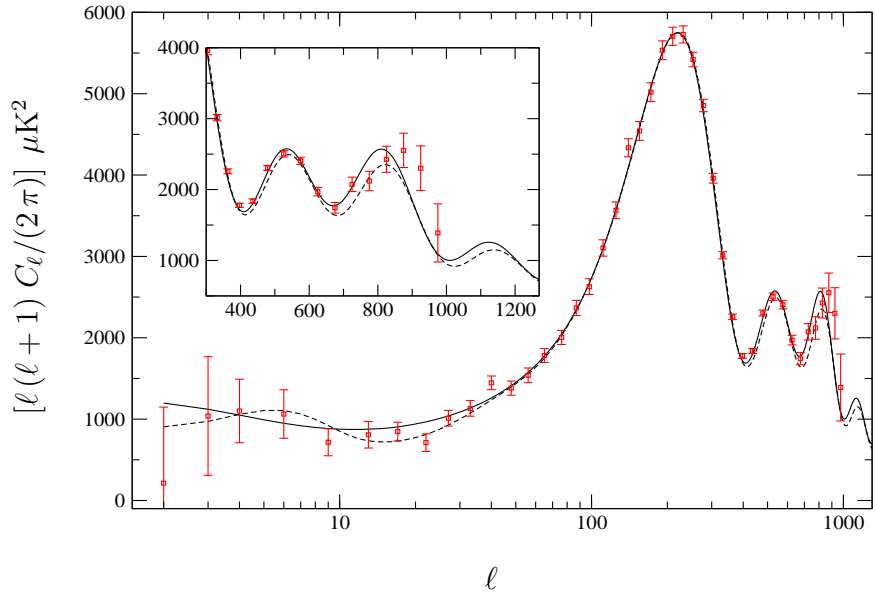}}
\vskip -165 true pt \hskip -340 true pt 
\rotatebox{90}{$\l[\ell\, (\ell+1)\;
C_{\ell}/(2\,\pi)\r]\; \mu{\rm K}^{2}$}
\vskip 50 true pt \hskip 10 true pt $\ell$
\vskip 2 true pt
\end{center}
\caption{The CMB angular power spectrum for the best fit values of 
the $n=3$ case (dashed line) and the best fit power law, reference
model (solid line) (cf.~Tab.~\ref{tab:bfv}).
Visually, it is evident that our model fits the data much better 
than the standard power law case at the lower multipoles.
The inset highlights the difference between our model and the power
law spectrum at the higher multipoles.
This difference arises due to the fact that, while the spectral index 
in the power law case is about $n_s \simeq 0.955$, the asymptotic 
spectral index in our case turns out to be $n_s \simeq 0.945$.}
\label{fig:cl}
\end{figure}

In Fig.~\ref{fig:cl}, we have plotted the angular power spectrum 
of the CMB temperature anisotropies for the best fit values of the 
parameters for the $n=3$ case.
For comparison, we have also plotted the angular power spectrum for 
the best fit reference model. 
It is immediately obvious from the figure that our model fits the 
lower multipoles much better than the reference model. 
As we have mentioned above, we obtain an improvement in 
$\chi_{\rm eff}^{2}$ of $6.62$ at the cost of introducing just one 
additional parameter when compared to the standard power law case.
We should also emphasize here that the improvement in the fit that 
we have achieved is not only due to the cut-off in the scalar power 
spectrum, but also because of the presence of the oscillations at the 
top of the spectrum, just before it turns scale invariant. 
Also, note the difference in the angular power spectrum for our 
model and the standard power law spectrum at the higher multipoles, 
which we have illustrated in the inset in Fig.~\ref{fig:cl}.
This disparity essentially arises due to the difference in the
asymptotic spectral index in our model (which proves to be about
$n_s \simeq 0.945$) and the spectral index in the power law case 
(which is about $n_s \simeq 0.955$, cf.~Tab.~\ref{tab:bfv}). 
The PLANCK mission~\cite{planck} is expected to provide more accurate 
data at these higher multipoles and, therefore, may aid us discriminate
between these models better.


\section{Summary and discussion}\label{sec:discussion}

In this section, after a quick summary of our results, we compare the
results we obtain with those obtained in another single scalar field 
model that has been considered earlier.
We emphasize the fact that the difference between these models has 
immediate observational consequences.


\subsection{Summary}

In this work, we have investigated a two stage slow roll 
inflationary scenario sandwiching an intermediate period 
of deviation from inflation, driven by potentials that are similar in 
shape to certain MSSM potentials~\cite{allahverdi-2007}.
In the MSSM case, inflation occurs when the field values are much 
smaller than the Planck scale~\cite{allahverdi-2007,sanchez-2007,allahverdi-2008}. 
However, in our case, since we demand two epochs of slow roll, it
necessarily requires that the initial values of the field (assuming, 
say, about $60$ $e$-folds of inflation) be greater than $M_{_{\rm Pl}}$.
The period of fast roll period produces a sharp drop in the scalar power 
spectrum for the modes that leave the Hubble radius just before the 
second slow roll phase.
We choose our scales such that the drop in power corresponds to the
largest cosmological scales observable today.
We find that the resulting scalar power spectrum provides a much better 
fit to the recent WMAP data than the canonical, nearly scale invariant, 
power law, primordial spectrum.


\subsection{Discussion}

At this stage, it is important that we compare our results with 
those obtained in another single field model that has been studied 
before.
As we had mentioned in the introduction, an initial kinetic 
dominated (i.e. fast roll) stage preceding slow roll inflation, 
driven by a quadratic potential has been considered earlier 
to provide a sharp drop in the scalar power spectrum at large
scales~\cite{contaldi-2003,nicholson-2008}.
At first glance, one may be tempted to conclude that the model we 
have studied here is equivalent to such a scenario if we disregard 
the first slow roll stage, since we have a kinetic dominated phase 
preceding a period of slow roll inflation. 
However, there are crucial differences between the two models which 
we have outlined below.

To begin with, in the model we consider, there is no freedom to 
choose the type of fast roll (say, the equation of state during 
the epoch of fast roll).
It is fixed once we have chosen the parameters so as to fit the 
observations. 
Secondly, in the scenario considered earlier, the modes which 
are outside the Hubble radius during the kinetic dominated 
phase {\it would have always remained so}\/ in the 
past~\cite{contaldi-2003,nicholson-2008}. 
The authors assume that somehow there may have been a previous 
phase of inflation, during which they were inside the Hubble 
radius and began life in the Bunch-Davies vacuum. 
While it is not impossible to think of situations where there may have 
been a previous inflationary epoch---for instance, it can be achieved 
by invoking another scalar field~\cite{polarski-1992-95,langlois-1999,tsujikawa-2003}---the consequences 
can be quite different.  
In contrast, in the scenario that we have considered here, the  
standard, sub-Hubble, Bunch-Davies, initial conditions have been
imposed on {\it all}\/ the modes. 
Thirdly, it was argued that, since the suppression of power for 
the scalar spectrum proves to be sharper than that of the tensor, the 
tensor-to-scalar ratio~$r$ displays a sharp rise towards large 
physical scales~\cite{nicholson-2008}, a feature that may 
possibly be detected by upcoming missions such as, for instance, 
PLANCK~\cite{planck}. 
However, in the models that we consider, the tensor amplitude 
on scales of cosmological interest proves to be too small 
($r <10^{-4}$) to be detectable in the very near future.
In conclusion, we would like to mention that a detection of the 
$C_{\ell}^{^{\rm BB}}$ modes corresponding to, say, $r > 10^{-4}$, 
can rule out the class of models that we have considered in this 
work.
 

\subsection*{Acknowledgments}

RKJ and LS would like to thank the Inter-University Centre for 
Astronomy and Astrophysics (IUCAA), Pune, India, and the Korea 
Institute for Advanced Study (KIAS), Seoul, Korea, for 
hospitality, where part of this work was carried out.
We would also like to acknowledge the use of the high performance 
computing facilities at the Harish-Chandra Research Institute, 
Allahabad, India, as well as at IUCAA and KIAS.
We wish to thank Anupam Mazumdar, Biswarup Mukhopadhyaya and Arman 
Shafieloo for discussions.
We also wish to thank Hiranya Peiris and Alexei Starobinsky for 
valuable comments on our results. 
We would also like to thank Tuhin Ghosh, Minu Joy, Juhan Kim, Subharthi 
Ray and Arman Shafieloo for help with the numerical codes at various 
stages.
JG is partly supported by the Korea Research Foundation Grant 
KRF-2007-357-C00014 funded by the Korean Government.
Finally, we acknowledge the use of the COSMOMC package~\cite{cosmomc}
and the data products provided by the WMAP science team~\cite{lambda}.


\end{document}